\documentstyle[12pt,aaspp4]{article}
\def\etal{{et al.}}
\def\asca{{\it ASCA}}
\def\iue{{\it IUE}}
\def\ex{{\it EXOSAT}}

\def\xte{{\it RXTE}}

\def\ros{{\it ROSAT}}

\begin{document}

\title {NEW CONSTRAINTS ON THE CONTINUUM-EMISSION MECHANISM OF AGN:
INTENSIVE MONITORING OF NGC 7469 IN THE X-RAY AND ULTRAVIOLET}

\author {K. Nandra\altaffilmark{1, 2}, 
J. Clavel\altaffilmark{3}, 
R.A. Edelson\altaffilmark{4},
I.M. George\altaffilmark{1,5},
M.A. Malkan\altaffilmark{6},
R.F. Mushotzky\altaffilmark{1},
B.M. Peterson\altaffilmark{7} and
T.J. Turner\altaffilmark{1,5}
}

\altaffiltext{1}{Laboratory for High Energy Astrophysics, Code 660, 
	NASA/Goddard Space Flight Center,
  	Greenbelt, MD 20771}
\altaffiltext{2}{NAS/NRC Research Associate}
\altaffiltext{3}{ISO Observatory, European Space Agency. Apartado 50727,
28080 Madrid, Spain}
\altaffiltext{4}{X-ray Astronomy Group, Department of Physics and
Astronomy, University of Leicester, University Road, Leicester,
LE1 7RH, United Kingdom}
\altaffiltext{5}{Universities Space Research Association}
\altaffiltext{6}{Department of Astronomy, University of California, 
Los Angeles,CA 90024 }
\altaffiltext{7}{Department of Astronomy, The Ohio State University,
174 West 18th Avenue, Columbus, OH 43210}

\slugcomment{Resubmitted for publication in
{\em The Astrophysical Journal}}

\begin{abstract}

We have undertaken near-continuous monitoring of the Seyfert 1 galaxy
NGC 7469 in the X-ray with \xte\ over a $\sim 30$~d baseline. The
source shows strong variability with a root-mean-square (rms)
amplitude of $\sim 16$~per cent, and peak--to--peak variations of a
factor of order 2.  Simultaneous data over this period were obtained
in the ultraviolet (UV) using \iue, making this the most intensive
X-ray UV/X-ray variability campaign performed for any active galaxy.

Comparison of the continuum light curves reveals very similar
amplitudes of variability, but different variability characteristics,
with the X--rays showing much more rapid variations.  The data are not
strongly correlated at zero lag. The largest absolute value of the
correlation coefficient occurs for an anticorrelation between the two
bands, with the X-ray variations leading the UV by $\sim 4$~d.  The
largest positive correlation is for the ultraviolet to lead the X-rays
by $\sim 4$~d. Neither option appears to be compatible with any simple
interband transfer function.  The peak positive correlation at $\sim
4$d occurs because the more prominent peaks in the UV light curve
appear to lead those in the X-rays by this amount. However, the minima
of the light curves are near-simultaneous.  These observations provide
new constraints on theoretical models of the central regions of active
galactic nuclei.  Models in which the observed UV emission is produced
solely by re-radiation of absorber X-rays are ruled out by our data,
as are those in which the X-rays are produced solely by Compton
upscattering of the observed UV component by a constant distribution
of particles.  New or more complex models must be sought to explain
the data.  We require at least two variability mechanisms, which have
no simple relationship.  We briefly explore means by which these
observations could be reconciled with theoretical models.

\end{abstract}


\section{INTRODUCTION}
\label{Sec:Introduction}

The origin of the continuum emission of Active Galactic Nuclei (AGN)
-- which covers an extremely broad band -- is not well understood.  In
a number of high luminosity sources, it appears that this emission
peaks in the ultraviolet (UV), the so-called ``Big Blue Bump''
(Shields 1978; Malkan \& Sargent 1982).  Strong, and apparently
non-thermal X-ray flux is also a persistent property of AGN (e.g.,
Marshall \etal\ 1981). The X-ray emission covers a wide band from at
least 0.1-100~keV, and can be described by a power-law form 
(Mushotzky, Done \& Pounds 1993).

The Big Blue Bump is often identified as the thermal output of an
accretion disk (henceforth the accretion disk model e.g., Shakura \&
Sunyaev 1973). Heat is generated by viscous dissipation in the disk,
which then radiates in the optical/UV regime for black hole masses
typical of AGN (e.g., Sun \& Malkan 1989).  An alternative origin for
the UV continuum emissions has been suggested by both observational
and theoretical considerations. Guilbert \& Rees (1988) postulated
that the UV need not be internally generated in the accretion disk,
but could arise via absorption and thermal re-emission (hereafter
referred to as ``thermal reprocessing'') of X-rays in optically thick
gas close to the central engine.  The material - which could be but
does not necessarily have to be the disk -- would imprint features on
the X-ray spectra (e.g., Lightman \& White 1988; George \& Fabian
1991; Matt, Perola \& Piro 1991). Such features have been found (e.g.,
Nandra \& Pounds 1994) and suggest that approximately half of the
incident X-rays are absorbed in the optically thick material.
Spectroscopic observations of strong gravitational and Doppler effects
in the iron K$\alpha$ line profiles of Seyfert galaxies (e.g., Tanaka
\etal\ 1995; Nandra \etal\ 1997) suggest that this material lies
extremely close to the central black hole and is probably in the form
of a disk (Fabian \etal\ 1995).  The bulk of the continuum photons
absorbed in the gas should then be re-emitted at the characteristic
thermal temperature of the material. For dense gas close to the
central engine, and particularly for  ``standard'' accretion disks, this
should be in the optical/UV.

Most models for the X--ray continuum of AGN are based on the idea that
lower-energy photons are Compton scattered by a population of hot
electrons and/or pairs (which we refer to as ``upscattering'' models
e.g., Sunyaev \& Titarchuk 1980; Svensson 1983; Guilbert, Fabian \&
Rees 1983). The seed photons are often assumed to be those in the blue
bump.  Specific models differ primarily in their assumptions about,
e.g., the geometry of the system (e.g. Haardt \& Maraschi 1991, 1993;
Haardt, Maraschi \& Ghisellini 1994; Stern \etal\ 1995), the question
of whether the electron population has a thermal or non-thermal
distribution, the importance of pairs (e.g., Zdziarski \etal\ 1990,
1994).

These models have been successful in explaining various observations.
The goal underlying the exploration of the models is the discovery of
the process responsible for the generation of the copious energy output
of AGN.  While the case for accreting supermassive black holes is
becoming compelling, the method by which the rest-mass energy of the
material is converted into radiative energy is still highly uncertain.
Some specific questions which remain about the emission mechanisms
include:

\begin{enumerate}
\item{How important is viscous dissipation in the generation of the UV?}
\item{What proportion of the UV arises via thermal reprocessing of X--rays?}
\item{What is the seed population for upscattering into the X--rays?}
\item{What mechanism accelerates the particles which up-scatter these 
seed photons?}
\end{enumerate}

A powerful way of investigating these questions is by variability
campaigns. These have already reaped rich rewards in the study of AGN
emission lines via ``reverberation mapping'' (e.g. Peterson 1993,
Netzer \& Peterson 1997, and references therein). These emission line
campaigns, however, also had strong implications for the generation of
the continua, which we shall discuss below. The models discussed above
all imply strong connections between the continuum emission in
different bands.  For example, the accretion disk emission could cover
an extremely broad band, depending on the temperature profile of the
disk. The thermal reprocessing model predicts that the X-rays should
be generating UV emission. The upscattering model suggests the
converse. By observing the variability in these bands, therefore, we
can make inferences as to which of the various processes is in
operation and to what degree. In particular, simultaneous X-ray/UV
data should be the most revealing.

A number of AGN have been monitored simultaneously at optical/UV and
X-ray energies.  Leaving aside blazars, the best-studied sources are
NGC~4151, NGC~5548 and NGC 4051. In the first two objects, there is
evidence for a correlation between the two bands. The best-sampled
(and therefore most reliable) case is NGC~4151, in which the 1455~\AA\
and 2--10~keV flux appears to correlate well on all time scales from
hours to a year (Perola \etal\ 1986; Edelson \etal\ 1996).  In
NGC~5548, the flux in the two bands is also well correlated on time
scales from days to 1 year (Clavel \etal\ 1992). In both sources
however, the correlation appears to break down during one very large
UV outburst.  NGC~4051 shows different behavior, in that the X-ray
emission showed large-amplitude (factor $\sim 2$) variability, while
the optical emission remained steady to within a few per cent when
observed over a $\sim 2$~d baseline (Done \etal\ 1990).  For
completeness, we also mention the results obtained for other
non-blazar AGN, though their significance is marginal due to the small
number of simultaneous observations and/or the short duration of the
campaigns.  In Fairall~9, the slow decline of the 2--10~keV flux
mimics the secular fading of the UV and optical continuum from 1978 to
1985 (Morini \etal\ 1986).  The UV-optical versus X-ray flux
correlation seems to hold in NGC~4593 (Santos-Ll\'{e}o \etal\ 1995),
whereas in MCG-8-11-11 (Treves \etal\ 1990), 3C120 (Maraschi \etal\
1991) and 3C~273 (Courvoisier \etal\ 1990) the two wavebands appear to
be independent of each other.  These previous attempts at determining
the relationship between the components have obviously left some
ambiguity. This is perhaps not surprising as generally the sampling of
the light curves has been rather poor.

In order to provide an improved dataset, a campaign of near-continuous
\xte\ and \iue\ monitoring of NGC 7469 was undertaken over a $\sim
1$~month baseline. The results of the campaign in terms of the
relationship of the X-ray and UV variability are the subject of this
paper. We have effectively divided the paper into two halves. Sections
2-4 discuss the observational results exclusively, which are then
summarized in Section 5. Section 6 then investigates the implications
of the observational results within the framework of the models
discussed above and suggest possible ways of reconciling the data with
models. 

\section{OBSERVATIONS}

Full details of the \iue\ data analysis and reduction, and the nature of
the UV variability has been described in Wanders \etal\ (1997;
hereafter W97).  Here we provide details of the X-ray observations,
made with the \xte\ Proportional Counter Array (PCA). The \xte/HEXTE
data will be presented elsewhere.

NGC 7469 was observed by \xte\ for a $\sim 30$~d period beginning 1996
June 10 02:26:40 UT. The nominal observational strategy was to observe
once-per \xte\ orbit ($\sim 90$~min) with a 1.5 ks exposure.  In
practice the exposure time in each orbit varied, with some orbits
containing no useful data, and others with up to 4ks of good data.
Approximately 10d into the campaign operations switched to one
observation every other orbit due to satellite power constraints.  We
utilized the so-called {\tt STANDARD-2} data mode which provides a
minimum time resolution of 16s, full detector and layer identification
and full spectral information. Due to temperature fluctuations and
arcing, 2 of the 5 proportional counter units (PCU's) which make up
the PCA were occasionally turned off during the campaign. For
simplicity and consistency here we only consider data for those three
PCU's (PCU0, PCU1, PCU2) which were operating throughout, and quote
all count rates for the sum of these three detectors.  Some details of
the \xte\ PCA in-orbit performance can be found in Jahoda \etal\
(1996).

\subsection{Data reduction}

Initial data reduction proceeded on the {\tt STANDARD-2} event files
by creating time filters based on various housekeeping
parameters. These are listed in Table~\ref{tab:sel}. Briefly, these
criteria were used to reject earth-occulted data and data contaminated
by the earth's limb, to exclude periods when the pointing position of
the satellite was unacceptable, and to reject anomalous background
events due to electrons. After applying these criteria we
combined all the {\tt STANDARD-2} data into a single light curve,
restricting the pulse-height range to obtain only the counts in the
2-10~keV band of the first of the three layers, where the PCA is most
sensitive.  We obtained $593$~ks of acceptable data in total.

\subsection{Background subtraction}

Background subtraction is the most important consideration when analyzing
\xte\ data for relatively weak sources such as NGC 7469.  The PCA has
no simultaneous background monitoring and therefore a model background
must be constructed based on various housekeeping parameters.  The
\xte\ background model is still evolving, but the current
implementation consists of three components: particle-induced
background, activation in the South Atlantic Anomaly (SAA), and the
diffuse X-ray background.

The particle-induced components are modeled by analyzing earth data
from throughout the mission. The main particle component is found to
be well correlated with many of the PCA anti-coincidence rates, which
are used to flag and veto non X-ray events. Many such rates are
available and the current model uses the so-called ``Q6'' rate, which
measures coincidences which fire 6 of the 8 PCU anode chains. In
addition to this primary particle component, the satellite's passage
through the SAA induces radioactive decay terms into the
background. These can be accounted for by correlating the positional
history of the satellite with the PCA count rate during occultation.
The sky background is determined by subtracting the predicted particle
components from ``blank'' sky pointings, averaging over several
positions. The mean background rate predicted for our observation was
$10.79$~ct s$^{-1}$ for the top layers of the three detectors in the
2-10 keV band.

\begin{figure}
\plotone{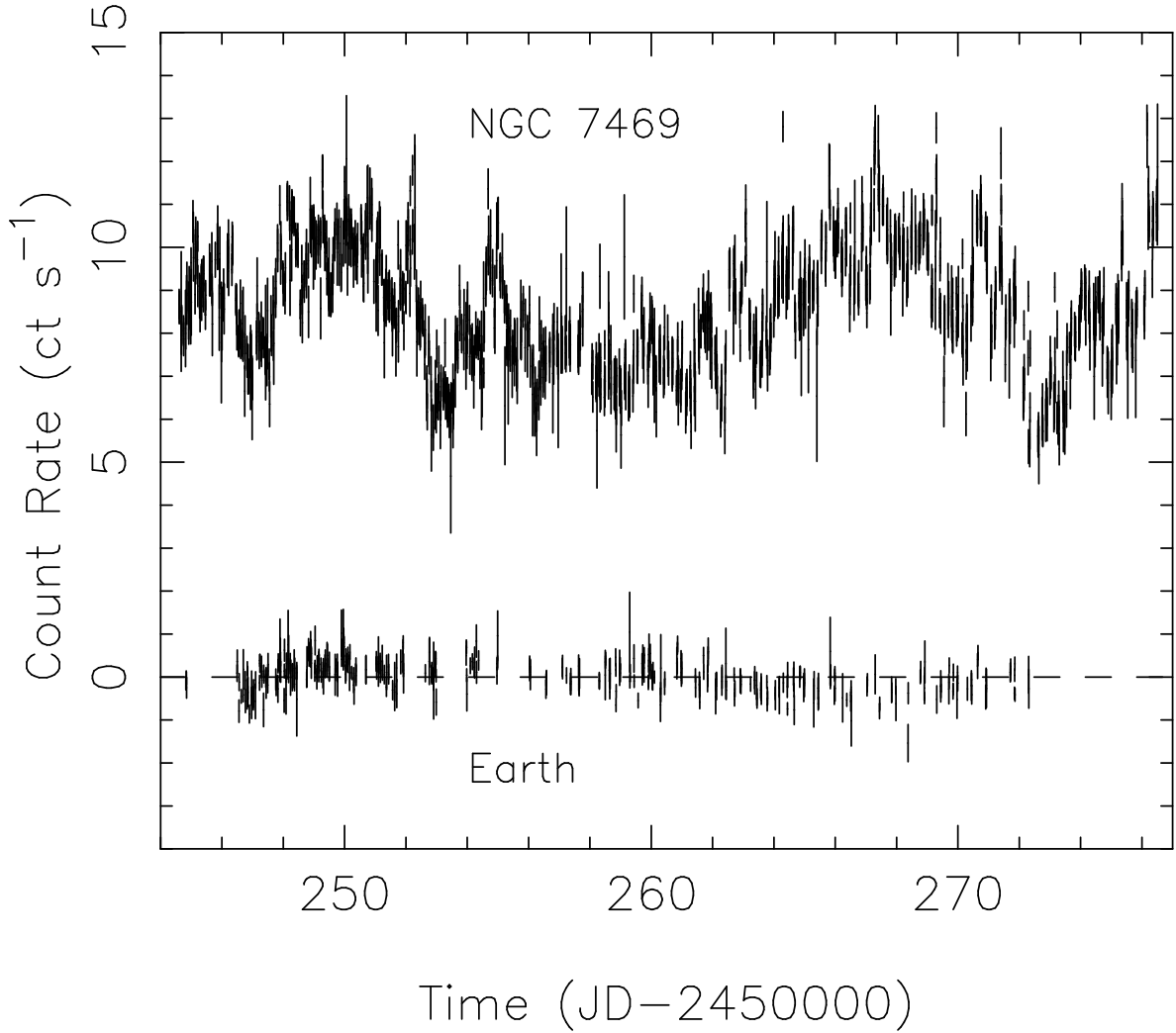}
\caption{
X-ray light curve (\xte, 3 PCUs, 2-10 keV), after background
subtraction, binned at 512s time resolution (upper points). The lower
points show the background-subtracted light curve of the
earth-occulted data. These data provide an estimate of the quality of
the subtraction of the particle background for the \xte/PCA. The
dashed lines shows the zero count rate level, with which the
earth-occulted data should be consistent after background
subtraction.  Those data show only small-amplitude variations (on
relatively long time scales) which show clearly that the variations in
the light curve on NGC 7469 are real, and not an artifact of poor
background subtraction.
\label{fig:sce_bgd}}
\end{figure}

As our interest is in the X-ray variability of the source in the 2-10
keV band, point-to-point fluctuations in the diffuse X-ray background
-- being constant in time -- will not affect our conclusions regarding
the variability.  More serious are potential inaccuracies in the
modeling of the particle background which is variable on a wide range
of (potentially characteristic) time scales. Therefore, to test the
accuracy of the background model at our epoch we accumulated the
earth-occulted data obtained during our campaign and applied the
particle background model. This analysis allows us to test systematic
uncertainties in the background subtraction, as with a perfect
predictive model the mean of this light curve should be consistent
with zero, and the variance in the data should
be due only to photon counting statistics. We did find variance in
this light curve, however, which we attribute to imperfect modeling of
the background. The magnitude of the excess variance depends on the
binning time scale. Fig.~\ref{fig:sce_bgd} shows the
background-subtracted source and earth-occulted light curves, with
a bin size of 512s.  Note that no earth-occulted datum is plotted after
day 272 as after this time, large
flares are apparent ($\sim 60$~ct s$^{-1}$ peak). 
The origin of this behavior is
unknown, but a detailed examination of these periods reveals that the
count rate begins to flare {\it only} when the Earth elevation angle
becomes less than zero, which strongly indicates that they are a
phenomenon associated with pointing at the earth. The
background-subtracted source data do not show such flares, and the
housekeeping data do not show unusual values during this time period.
We therefore conclude that this phenomenon does not affect our light
curve of NGC 7469.

Even when these data are excluded, the background-subtracted light
curve of the earth-occulted data shows long term trends, which are
currently not accounted for in the background model.  Excluding the
flaring periods, we obtained a mean value of $0.032 \pm 0.010$ ct
s$^{-1}$ and estimate a variability amplitude of $0.24$~ct s$^{-1}$
(1$\sigma$) in the background with a time binning of 512s.  This
systematic error was added in quadrature to the statistical error for
each data point in the source light curve. In spite of the systematic
nature of these errors, which makes them difficult to propagate
correctly, the variations in the source-minus-background light curve
are so large that we conclude that they are intrinsic to NGC
7469. This is amply demonstrated by comparing the excess variability
in the earth-occulted data to the source variations in the same energy
band of $1.38$~ct s$^{-1}$, a factor $\sim 6$ larger, and an
examination of Fig.~\ref{fig:sce_bgd}.

\section{INTERBAND CONTINUUM RELATIONSHIPS}

\begin{figure}
\plotone{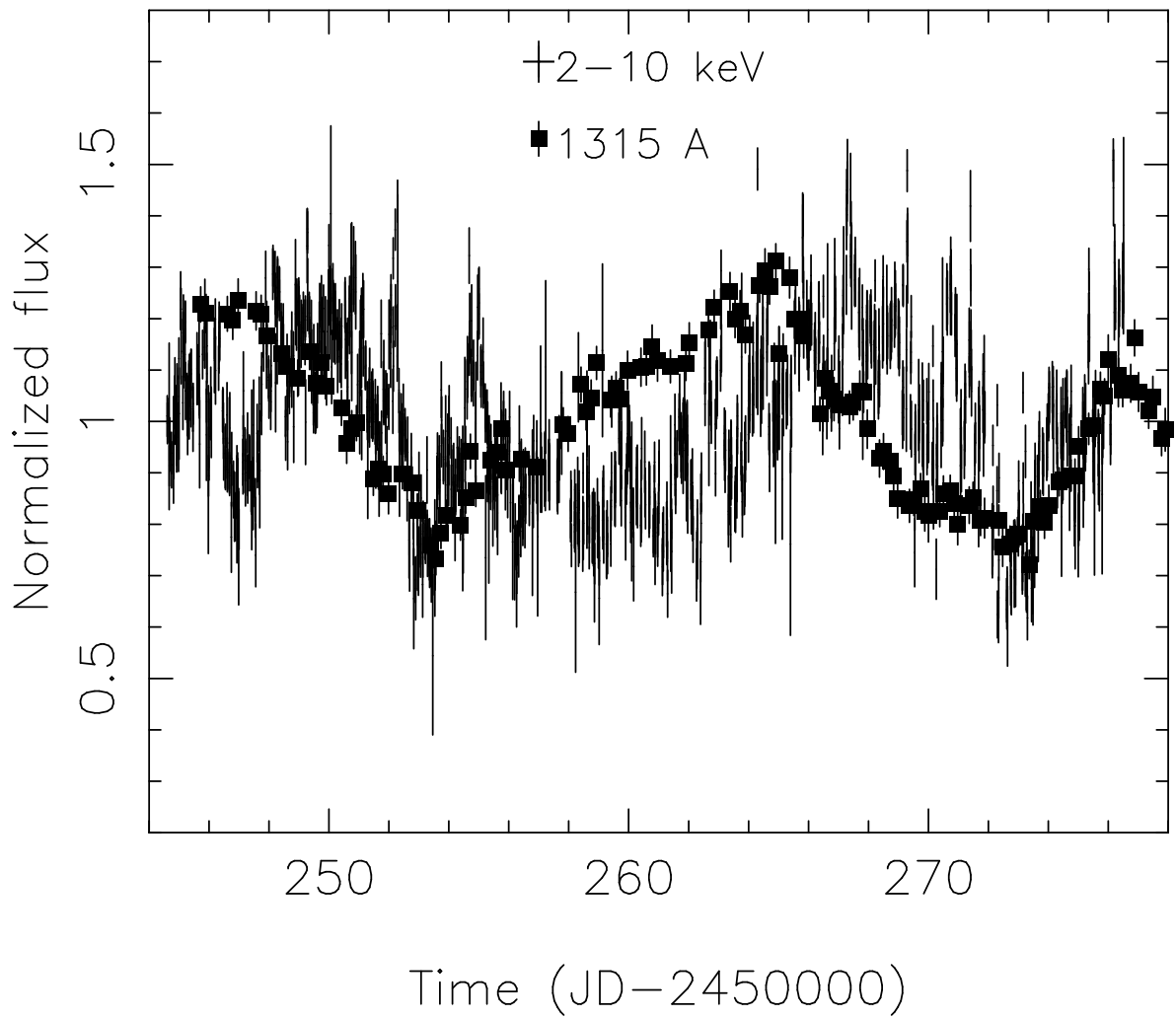}
\caption{X-ray light curve (\xte\ PCA, 2-10 keV) binned
at 512s time resolution (crosses) together with the \iue\ 1315\AA\
light curve in the UV (squares; W97).
Both light curves have been normalized to their mean value and
are plotted on the same time axis. We note the remarkable similarity
in the amplitudes of variability on long time scales, but clear
evidence for more rapid X-ray variations which are not observed in the UV.
The light curves are not well correlated at zero lag, but the two prominent
troughs appear to line up rather well in the time domain. In contrast,
the apparent peaks in the UV light curve appear to precede the
X-ray peaks. 
\label{fig:lc}}
\end{figure}

After background subtraction, the mean source count rate was found to
be $8.67 \pm 0.01$~ct s$^{-1}$ in the 2-10 keV band (statistical error
only), which corresponds to a flux of $\sim 3.4 \times 10^{-11}$\,erg
cm$^{-2}$ s$^{-1}$ assuming the \asca\ spectrum for this source given
by George \etal\ (1998). The corresponding 2-10 keV luminosity is
$4.2\times 10^{43}$\,erg s$^{-1}$ ($z=0.017$; we have assumed $H_{\rm
0}=50$ and $q_{\rm 0}=0.5$). Fig.~\ref{fig:lc} shows the
background-subtracted X-ray light curve of NGC 7469 in the 2-10 keV
band, in 512s bins, which has been normalized to this mean count
rate. There are clearly strong variations on many time scales, and we
find a fractional rms variability parameter $F_{\rm var}$ = 0.16,
defined as the square root of the excess variance of the data points,
divided by the mean (e.g., Rodriguez-Pascual et al., 1997).
Variability information is detailed in Table~\ref{tab:var}. This shows
the number of data points, $N_{\rm data}$ in each light curve, $F_{\rm
i} (t)$. The mean flux $\overline{F}$, defined as

$$
\overline{F} = \frac{1}{N_{\rm data}} \sum_{i=1}^{N_{\rm data}} F_{\rm i}(t)
$$

\noindent
The standard deviations of the points $\sigma_{\rm F}$ is also shown,
defined by:

$$
\sigma^{2}_{\rm F} = \frac{1}{N_{\rm data} - 1 } \sum_{i=1}^{N_{\rm data}}
(F_{\rm i}(t) - \overline{F})^{2}
$$

\noindent
The data points $F_{\rm i}(t)$ have associated error bars $\sigma_{i}(t)$
and we define the expected variance due to random errors as,

$$
\Delta^{2}_{\rm F} = \frac{1}{N_{\rm data}} \sum_{i=1}^{N_{\rm data}}
\sigma^{2}_{i}(t)
$$

\noindent
Then, we define the excess variance $\sigma_{\rm XS}$ as the
difference between the observed and expected variances, 

$$
\sigma^{2}_{\rm XS} = \sigma^{2}_{\rm F} - \Delta^{2}_{\rm F}
$$

\noindent
and the rms variability parameter as,

$$
F_{\rm var} = \sigma_{\rm XS}/\overline{F}
$$

From Fig.~\ref{fig:lc} it can be seen that the minimum-to-maximum
variations are roughly a factor $\sim 2$. Note that normalized count
rate has been plotted in Fig~\ref{fig:lc}, rather than the flux. This
can cause some additional uncertainty if the spectrum is strongly
variable. We defer a full discussion of the X-ray spectrum of NGC~7469
and its variability to a later paper, but a preliminary examination of
the hardness ratio suggests that there are no strong spectral changes
during our observation.

The normalized \iue\ continuum light curve at 1315\AA\ (W97) is also
shown in Fig.~\ref{fig:lc} (filled squares).  Several things are
apparent from a comparison of these two plots. The most striking
result of our simultaneous campaign is that the X-ray and UV fluxes
are not well correlated at zero lag.  We also note with interest that
the amplitude of variations in the X-ray and UV are very similar, with
$F_{\rm var}$ in the two bands of 0.16 and 0.15 respectively
(Table~\ref{tab:var}). These values are dominated by the events on
long time scales, which have the largest amplitude.  It is tempting to
match up the largest amplitude peaks and troughs in the two light
curves and conclude that there is a significant correlation, but with
a time-delay between the two bands.  The large-amplitude trends in the
light curves show two relatively well-defined maxima and minima.
However, the data are clearly inconsistent with a single or simple
time delay. The maxima show an apparent (and similar) delay of $\sim
4$~d between the UV and X-ray, but the minima appear to occur
simultaneously.  We investigate this further below.  There are also
some highly significant and very rapid events in the X-ray light curve
which are not seen with the same amplitude in the UV (e.g., the events
around days 252 and 255).

\subsection{Auto-correlation functions}

\begin{figure}
\plotone{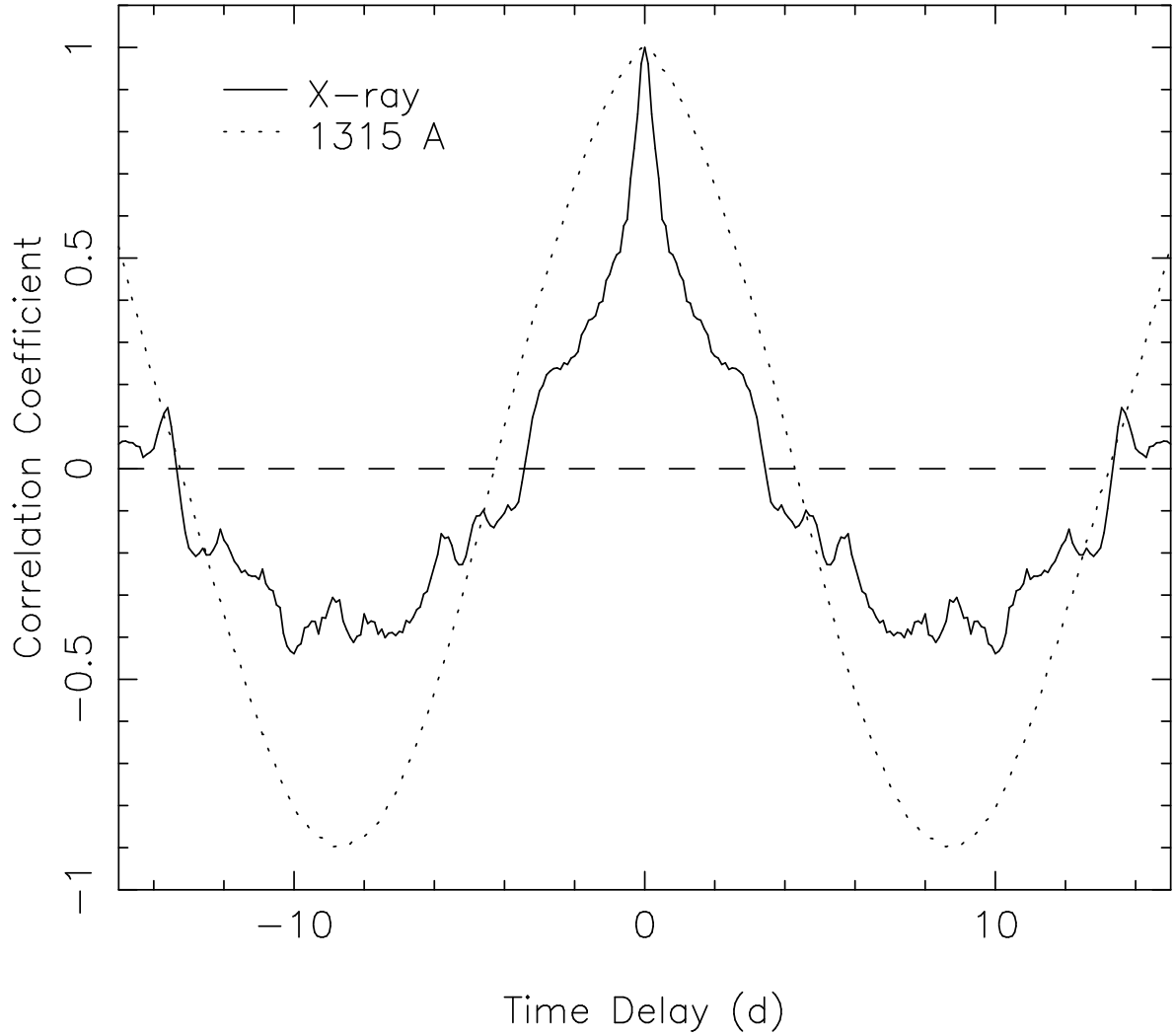}
\caption{
Auto-correlation functions of the X-ray (solid line) and
UV 1315\AA\ (dotted line) light curves. The X-ray ACF
is much narrower (FWHM$= 1.7$~d) than that in the
UV (FWHM $= 5.3$~d), illustrating the rapid
variability which is not observed in
the UV.
\label{fig:acf}}
\end{figure}

We have calculated the auto-correlation functions (ACFs) of the two
light curves and show these in Fig.~\ref{fig:acf}.  Before performing
these correlations, the UV light curve at 1315\AA\ was truncated to
cover the same time-span as the X-ray campaign.  Furthermore, to
provide the fairest comparison, we resampled the X-ray light curve to
have the same time resolution as the UV light curve. The X-ray
auto--correlation function is clearly narrower than that of the UV
continuum and has more structure. The Full-Width at Half Maximum
(FWHM) of the X-ray ACF is 1.7d, whereas the UV ACF has FWHM more than
twice this (5.3d).  This shows that the X-rays have more variability
on short time scales. The ACFs are subject to some uncertainty due to
the length of the data-train, but we feel this is unlikely to be a major
problem in our analysis. For example, if we compare the ACF of the
full UV light curve presented by W97 (which lasts $\sim 45$d) to the
truncated version, we obtain very similar values for the FWHM of the
UV ACF. We are therefore confident that the comparison of the FWHM is
not strongly affected by the sampling of the two light curves.

\subsection{Cross-correlation analysis}

\begin{figure}
%\epsscale{0.9}
\plotone{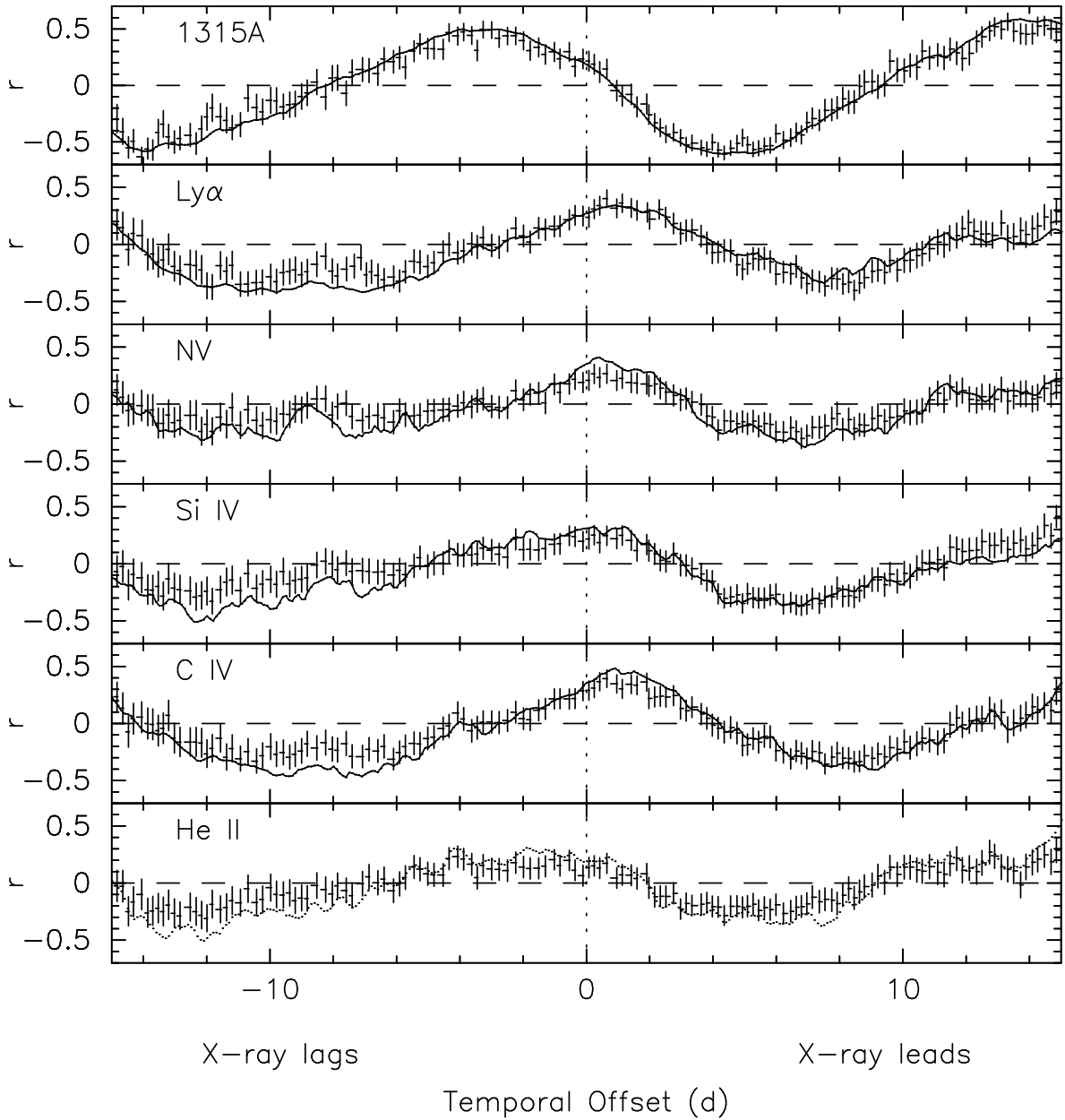}
\caption{
Cross-correlation functions for the X-ray light curve versus various
UV light curves. In all cases the crosses represent the ZDCF
(Alexander 1997) and the solid lines the ICCF (White \& Peterson
1994). The top panel shows the correlation of the 2-10 keV X-rays with
the continuum at 1315\AA. The zero-lag correlation is rather weak, but
a peak at a lag of $\sim -4$d and a trough at $\sim 4$d both have
apparently high statistical significances. The UV emission lines
(bottom five panels) are moderately well-correlated with the X-ray
continuum.  Indeed Ly$\alpha$, N{\sc v} and C{\sc iv} all correlate
better with the X-ray continuum than the 1315\AA\ continuum at zero
lag. However, if a lag is allowed, the UV continuum is more strongly
correlated.
\label{fig:xcor}}
\end{figure}

\begin{figure}
\plotone{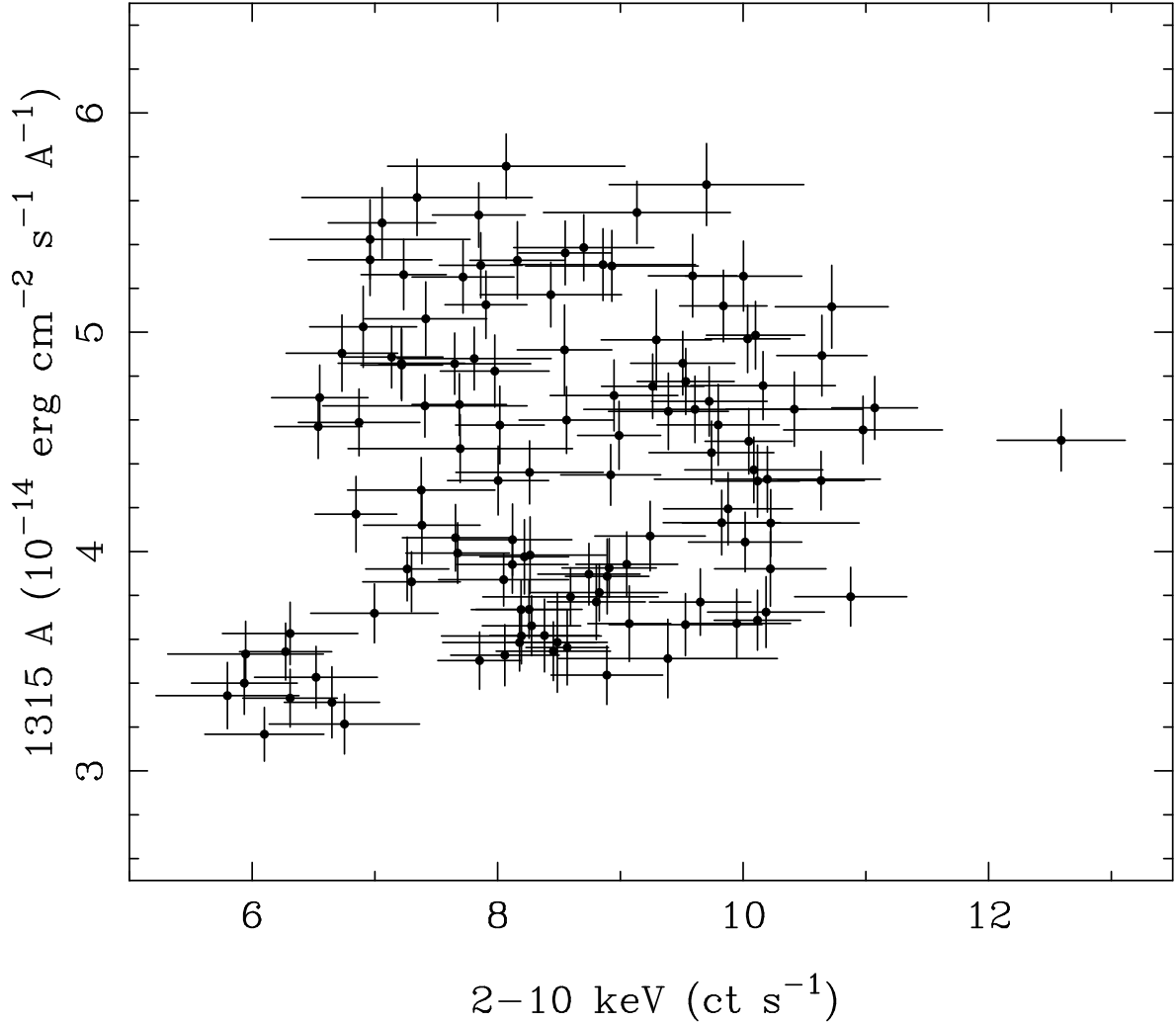}
\caption{
Estimated correlation of the X-ray count rate and UV flux.
Pseudo-simultaneous X-ray data were estimated by interpolating the
X-ray light curve to the times when the UV points were obtained. The
correlation coefficient at zero lag, estimated from the ZDCF
(Fig.~\ref{fig:xcor}) is found to be $r_{\rm 0} = 0.19$, which is
significant at 96~per cent confidence. The ICCF suggests a correlation
coefficient of $r_{\rm 0}=0.16$, which is not significant. The
correlation coefficient of the points shown is $r_{\rm 0}=0.15$
for 123 points, again not significant.
\label{fig:nolag_cor}}
\end{figure}

We have correlated the X-ray and UV continuum light curves using two
techniques for purposes of comparison, a cross-correlation with
interpolation (ICCF: Gaskell \& Peterson 1987; White \& Peterson
1994), and the z-transformed Discrete Correlation Function (ZDCF:
Edelson \& Krolik 1988; Alexander 1997).  We have evaluated the
correlation coefficient for delays in the range $-15<D<15$~d computing
the values in steps of 0.1~d for the ICCF.  The ZDCF binning is
described in Alexander (1997) but in practice was similar to the
ICCF. The results of this cross-correlation analysis are shown in the
top panel of Fig.~\ref{fig:xcor}. The two methods are in good
agreement. Fig.~\ref{fig:nolag_cor} shows the fluxes plotted against
one another, using a linear interpolation in the X-ray light curves to
estimate the X-ray flux at the time of the UV observations.  These
demonstrate the conclusion which is already apparent by eye, that the
light curves are not well-correlated at zero lag.  Using the ZDCF, and
a linear fit to nearby points, we estimate a correlation coefficient,
$r$, at zero lag of $r_{0}=0.19$ (see also Table~\ref{tab:ccf}).
Assuming that the pairs of X-ray/UV data points in the cross
correlation are independent, this implies a significant correlation at
$\sim 96$~per cent confidence.  A local fit to the ICCF near zero lag
gives a slightly smaller estimate for the correlation coefficient of
$r_{0}=0.16$.  This value of $r$ is not significant.  We conclude that
the two bands are not strongly correlated at zero lag.

Peaks and troughs are apparent in Fig.~\ref{fig:xcor}.  The most
significant value of $r$ is negative, implying an anti-correlation
between the bands when the X-ray variations lead those in the UV by $D
\sim 4$~d. A local Gaussian fit to the ZDCF shows a minimum
correlation coefficient of $r^{-}_{\rm max}=-0.57$.  Assuming the data
points are independent, the formal significance of obtaining this by
chance is $p_{\rm r} = 8 \times 10^{-9}$, after accounting for the
number of trials.  The most significant positive correlation when the
X-ray variations lag those in the UV by $D \sim 4$~d.  A Gaussian fit
to the ZDCF here gives a maximum value for the correlation coefficient
of $r^{+}_{\rm max} = 0.45$ (Table~\ref{tab:ccf}). After accounting
for the number of trials, the probability of obtaining such a
coefficient by chance is found to be $p_{\rm r} = 2\times 10^{-4}$,
assuming independent data points.  The ICCF gives very similar values.

On the face of it, it seems possible to conclude either that the
X-ray and UV light curves show a significant anti-correlation, with
the X-ray variations leading the UV by $D \sim 4$~d, or that two bands
are correlated, with the UV emission leading the X-rays by a similar
amount.  However, the interpretation of these possible lags is unclear
and we note that having sampled so few of the highest amplitude
``events'' with our baseline, aliasing problems are possible.  Also,
as already noted above, the character of the two light curves shown in
Fig.~\ref{fig:lc} shows that it is highly unlikely that the data would
be consistent with a single lag.

If one component were driving the other, a more complex relationship
between the two bands may be expected, depending on factors such as
the geometry of the system. In optical/UV reverberation studies, this
relationship is characterized by the ``transfer function'' (Blandford
\& McKee 1982). Once again, however, it seems difficult to reconcile
our data with a single transfer function.  In any simple and/or
constant geometry (and therefore transfer function) the variations in
the secondary band should lag those of the driver with a delay which
was similar for all ``events''.

\subsection{Power-density spectra}

Besides measuring the auto-correlation functions and interband
cross-correlation function, as has been done in the previous sections,
the statistical characteristics of the UV and X-ray variations can be
compared by measuring the fluctuation power density spectra (PDS) in
the different bands. This is made more powerful in this case because
it is the first time that simultaneous, quasi-continuously sampled
data have been obtained for any AGN on such long time scales. This
means that the PDS can be compared directly without having to worry
that the source was varying differently at different times, or about
the effects of different sampling patterns.  This is also the first
measurement with relatively evenly-sampled data of the X-ray PDS of a
Seyfert galaxy on time scales longer than a few days.

\begin{figure}
\plotone{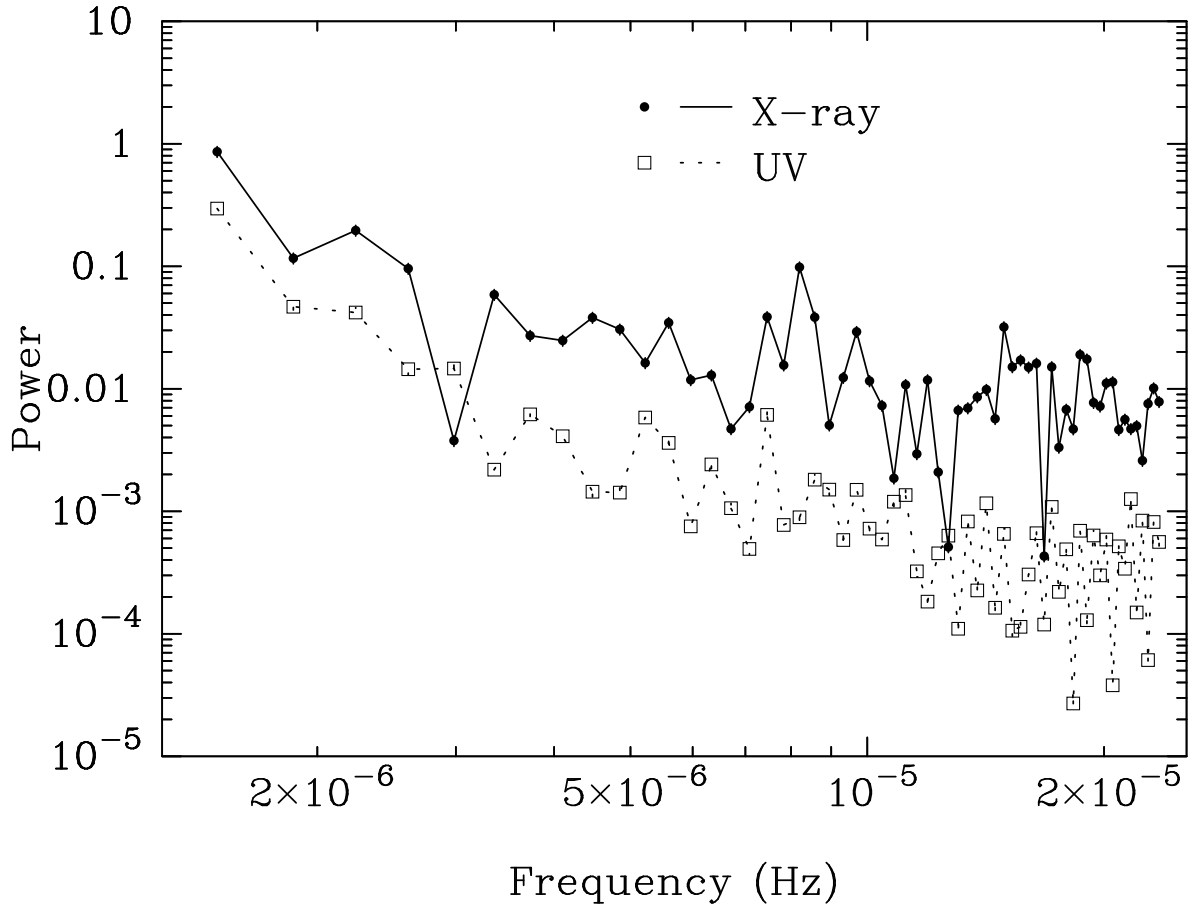}
\caption{
Periodograms for the X-ray (solid circles, solid line) and UV (open
squares, dotted line) light curves.  Both show a ``red-noise'' PDS
without any obvious characteristic variability time scale. At face,
value the power spectral slopes appear to be similar, although the
X-rays show larger amplitude variations. However, significant
statistical noise in the UV may make that PDS look artificially flat.
\label{fig:pds}}
\end{figure}

For this analysis, both the UV and X-ray light curves were
slightly resampled to yield 128 data points centered on identical
times.  The X-ray data were rebinned directly into 20947 sec
intervals. The rebinned UV light curve was constructed by
interpolating the measured light curve to the mid-point times used in
the rebinned X-ray light curve. Both light curves were
renormalized by subtracting the mean and dividing by the standard
deviation of the points. These data were then Fourier transformed with
a Parzen window (Press \etal\ 1992), with the results shown in
Fig.~\ref{fig:pds}.

As expected, both the UV and X-ray PDS show no narrow features that
would be indicative of periodic or quasi-periodic variability, but
instead the fluctuation power rises strongly to lower frequencies.
This ``red noise" PDS is a well-established feature of AGN (e.g.,
Krolik \etal\ 1991; Lawrence \& Papadakis 1993; Green, McHardy \&
Lehto 1993). \ex\ observations yielded PDS characterized over the
frequency range $10^{-3} <f< 10^{-5} $~Hz as $f^{-1.5}$ for most
Seyfert galaxies (Green \etal\ 1993; Lawrence \& Papadakis 1993),
although the slopes must flatten out to lower temporal frequencies or
the total power will diverge.  However, estimating the slopes for our
power spectra is complicated for two reasons. First, there are
insufficient points to re-bin the PDS and therefore assign error bars
to the points using the method of, e.g. Papadakis \& Lawrence
(1993). Second, the noise level due to statistical fluctuations,
particularly for the UV light curve, is difficult to
estimate. Regarding this latter point, we have performed simulations
of constant light curves, with the same sampling patterns as the
interpolated light curves used to produce the power spectra. Gaussian
noise was added to these assuming the mean error bars of the points to
estimate the distribution.  These were then Fourier transformed,
giving us an estimate of the noise level predicted by the error
bars. In the case of the X-ray PDS, we find a predicted noise level
which is a factor $\sim 2$ lower than the mean power at the high
frequency end of the PDS. Conversely, the UV noise level is a factor
$\sim 2$ {\it higher} than the observed power at high
frequencies. This suggests that the UV error bars are
overestimated. Indeed, W97 point out that their conservative method of
calculating the UV error bars will tend to produce upper limits.  A
good estimate of the slope of the UV PDS would require a more reliable
estimate of the noise level.

We note that the X-ray PDS clearly shows more power than the UV
across the whole range of temporal frequencies.  This is difficult to
reconcile with the fact the the $F_{\rm var}$ values, which should
represent the integral of the PDS over some frequency range, are so
similar. As seen in Fig.~\ref{fig:pds}, the X-ray PDS appears to be
only slightly flatter than the UV PDS. Naively, the observed
variability might lead one to predict that the X-ray PDS had a much
flatter slope, given clear eyeball evidence for more rapid
variability.  As noted above, statistical noise could have a strong
effect on the UV PDS, accounting for the bulk of the high frequency
power and making the PDS appear flatter.  However, these problems make
it difficult to make a meaningful statements regarding these
preliminary PDS.  

\section{CONTINUUM-LINE RELATIONSHIPS}

W97 have also presented the light curves of the UV emission lines
obtained during the \iue\ campaign, and performed cross correlations
of their fluxes with the 1315\AA\ continuum. As found for the original
campaign on NGC 5548 (Clavel \etal\ 1991; Peterson \etal\ 1991), the
line fluxes were found to lag those of the continuum by a few
days. Here we examine the relationship of the X-ray flux with that of
the UV emission lines. Table~\ref{tab:var} shows the variability
parameters for the emission lines.  Interestingly, we find that many
of the emission lines are reasonably well correlated with the X-ray
continuum close to zero lag. Fig.~\ref{fig:xcor} also shows the ZDCF
and ICCF correlation functions for the X-ray versus the prominent UV
emission lines and Table~\ref{tab:ccf} shows the correlation
coefficients.  Once again, we truncated the \iue\ light curves such
that they cover the same time as the X-ray data. We find that the
Ly$\alpha$, N {\sc v} and C {\sc iv} lines are all correlated better
with the X-ray continuum than the UV continuum at zero lag. However,
an examination of the peak correlation coefficients shows that if a
time delay is allowed, the UV continuum is better correlated with all
of the emission lines (Table~\ref{tab:ccf}).

\section{OBSERVATIONAL SUMMARY}

Here we summarize our observational results:

\begin{enumerate}
\item{The UV and X-ray continua of NGC 7469 varied by a factor $\sim 2$
peak-to-peak on time scales of $\sim 10$~d}
\item{The X-rays also showed much more rapid variations ($\sim 50$~per cent
in $\sim 1$~d, which were not observed in the UV}
\item{The variations in the two bands are not well correlated at
zero lag}
\item{The strongest peaks in the X-ray light curve are delayed by
$\sim 4$~d with respect to those in the UV}
\item{The prominent ``troughs'' in the light curves are
near-simultaneous}
\item{The highest absolute value of the correlation
coefficient occurs for an X-ray lead of $\sim 4$d and an
anticorrelation between the bands}
\item{The strongest positive correlation occurs for an
X-ray lag of $\sim 4$~d}
\item{Many of the prominent UV emission lines are better correlated
with the X-ray continuum at zero lag, but they are all better
correlated with the UV emission when a lag is allowed}
\end{enumerate}

\section{DISCUSSION}

We have investigated the relationship between the X-ray and UV
emission in NGC 7469 on time scales of hours-weeks.  The poor
correlation between the X-ray and UV light curves at zero lag may be
considered a surprising result because, as mentioned in \S1, some
previous experiments suggested a good correlation, and little if any
time lag between the variations in the two bands (e.g. NGC 5548, NGC
4151).  No other AGN, however, has been monitored as intensively and
for such a long duration.

Variability information has important implications for the physical
mechanisms responsible for the production of the X-ray and UV emission
in AGN. If the flux changes in two bands are correlated, this suggests
some causal link between them. A time lag between the bands then shows
which component drives the other. If X-ray variations lead those in
the UV, this is strongly suggestive of thermal reprocessing. If the
opposite is observed, this strongly favors upscattering. With a simple
transfer function this delay should be similar for all ``events'' in
the light curve. No such simple behavior was observed during our
campaign and the interpretation is less straightforward. Our data
require modifications to the simplest ideas about the emission
mechanisms in the UV and X-ray.

\subsection{Model ingredients}

It is widely accepted that the radiative energy from AGN originates as
the rest-mass energy of accreting material. The conversion process
must include a mechanism to accelerate the particles which produce the
X-rays as they are the highest-energy photons to carry an appreciable
fraction of the luminosity.  In principle, the remainder of the AGN
spectrum could be produced by thermal reprocessing following
absorption of some fraction of this X-ray continuum.  Viscous
dissipation in the accretion flow, however, could dominate the
observed UV emission.  In many scenarios there is also a radiative
connection between the UV and X-ray emission regions: X-rays can be
produced by upscattering of UV seed photons and UV emission can be
produced by re-radiation of absorbed X-rays. In these circumstances, a
substantial number of factors can affect the observed variability:

\begin{enumerate}
\item Changes in the physical properties (e.g. optical depth, temperature,
geometry) of the particle distribution which produces the X-rays
\item Instabilities in the accretion flow
\item The geometry and size of the X-ray and UV emission regions 
\item Anisotropy of the radiation fields, which might include
relativistic effects close to the black hole
\item The temperature distribution of the absorbing medium
\item The importance of feedback, in which variations in each 
band affect the other
\item Changes in occulting/absorbing media 
\end{enumerate}

\subsection{Implications of the NGC 7469 data}

Our data have a number of implications for the processes which produce
the UV--to--X-ray emissions of NGC 7469 and the observed variability,
which we now discuss. First we consider our observations in the
context of historical data and the spectral energy distribution of NGC
7469.

\subsubsection{The broad-band perspective}

The mean X-ray and UV fluxes observed for NGC 7469 during our campaign
are very much typical of the respective historical means for this
source. The mean flux in the 2--10 keV band, based on historical
observations over the period 1979-1993, is $\sim 3 \times
10^{-11}$\,erg cm$^{-2}$ s$^{-1}$. Furthermore, the range of
historical variability is very similar to that observed during our
campaign. This suggests that we sampled a large fraction of source
variability in NGC 7469 during our one-month campaign, although we
observe no obvious flattening of the PDS (Fig.~\ref{fig:pds}) at the
lowest frequencies.  Chapman, Geller and Huchra (1985) derived a mean
value of $4.6 \times 10^{-14}$\,erg~cm$^{-2}$~s$^{-1}$~\AA$^{-1}$ for
the 1430--1460~\AA\ continuum flux of NGC~7469 from 10 \iue\
observations in 1979--1982. Similarly, Edelson, Pike and Krolik
(1990), reported a mean value of $4.8 \times
10^{-14}$\,erg~cm$^{-2}$~s$^{-1}$~\AA$^{-1}$ for the continuum flux at
1450~\AA\ (rest wavelength) from 16 \iue\ observations in 1979--1985.
With a mean UV flux at 1485~\AA\ (observed wavelength) of $4.0 \times
10^{-14}$\,erg~cm$^{-2}$~s$^{-1}$~\AA$^{-1}$ (W97), NGC~7469 was thus
neither particularly faint nor exceptionally bright during our
campaign.  The optical--to--X-ray spectral index of NGC~7469,
$\alpha_{ox}$~=~1.22, is not significantly different from that of,
e.g., NGC~5548 ($\alpha_{ox}$~=~1.25), or the mean index for Seyfert
galaxies (Kriss, Canizares and Ricker 1980).

The spectral energy distribution of NGC 7469 is shown in
Fig.~\ref{fig:sed}. The only real peculiarity of NGC~7469 is the
presence of a circumnuclear starburst ring within $1.\!''5$ of its
nucleus. Genzel \etal\ (1995) estimate that the starburst accounts for
two-thirds of the source bolometric luminosity and it may dominate the
IR emission.  However, it only contributes 4 percent to the observed
X-ray flux (Perez-Olea and Colina 1996). The starburst should be
invariant on the time scales sampled here and have no effect on the
X--ray/UV variability.

\subsubsection{The X--ray continuum}

The presence of rapid variations in the X-rays which are not seen in
the UV implies that the particle distribution responsible for the
X-rays is variable.  The X-ray emission mechanism is highly uncertain,
but as mentioned in the introduction, many models have concentrated on
Compton upscattering of seed UV photons by a population of hot
electrons and/or pairs (e.g. Haardt \& Maraschi 1991, 1993). Our
observations show that if the observed UV photons are the seed
population, then the rapid variations of the X-rays do not arise from
variations in the seed.  Either the optical depth, temperature or
geometry of the upscattering region must be changing.  In the latter
case, changes in the distribution of active regions in the X-ray
source, or kinematic effects can produce variability (e.g., Abramowicz
\etal\ 1991; Haardt, Maraschi \& Ghisellini 1997).

Longer-term variations are also observed in the X-ray flux and these
could also arise from processes such as those just mentioned. The fact
that the power spectrum shows no obvious features or break is
supportive of this interpretation. The fluctuations have a similar
amplitude to those in the UV and this suggests a connection between
the bands.  This is intriguing. One possibility is that the long
time-scale variability in the X--rays is due to changes in the UV seed
population. In the simplest such interpretation - where the
upscattering region is point-like and lies in the line of sight to a
point-like seed source we expect a 1:1 correlation between the two
bands with no lag. This is ruled out by our data, although we note
that the minima appear to be very close in time. In more complex
geometries, there may be time lags which will be in the sense that the
X-ray variations follow those in the UV. The delay of $\sim 4$~d
between the peaks is superficial evidence in favor of upscattering.
In that model, however, we expect the lag to be very short, being
dependent primarily on the light travel time between the regions,
modified by geometrical factors. A lag of $\sim 4$~d, the only
plausible conclusion here, seems rather long to be associated with
these processes. In addition, we are unable to envisage a purely
geometrical modification of a linear process which accounts for {\it
both} the relationship between the maxima and that of the minima.

\subsubsection{The UV continuum}

\begin{figure}
\plotone{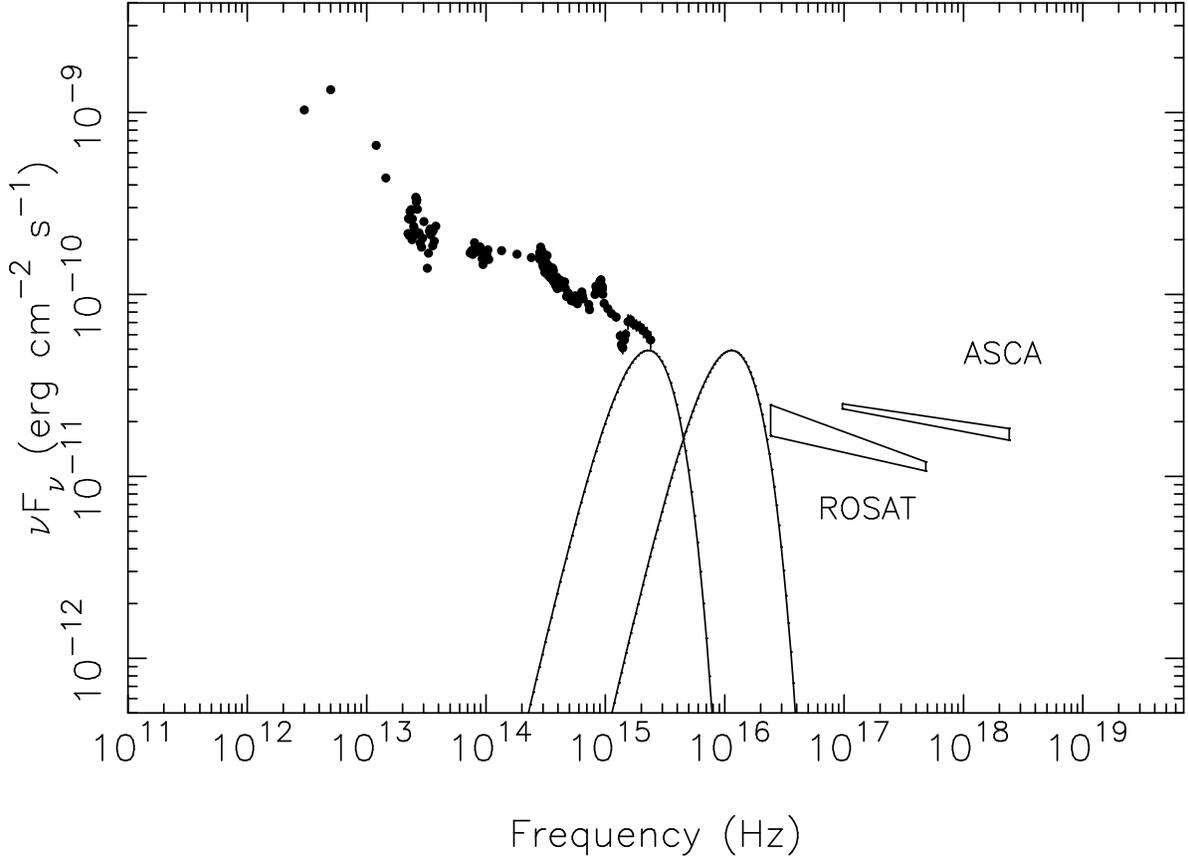}
\caption{
Spectral Energy Distribution for NGC 7469 from far-infrared to X-ray
frequencies. The data are not simultaneous. The UV-to-far IR data
were assembled from several instruments by Edelson \& Malkan (1986)
and references therein. The X-ray flux level
observed here with \xte\ was very similar to that seen by \asca
(Nandra \etal\ 1997; George \etal\ 1998). ROSAT data are from Brandt
\etal\ (1993). The UV emission is similar to the mean observed here.
The two solid lines shown are black body spectra with temperatures of
$2$ eV and 12 eV, and luminosity of $8 \times 10^{43}$~erg s$^{-1}$. This
represents our best estimate of the likely level of reprocessed X-ray
flux in NGC 7469. The cooler blackbody is optimized to peak at
1315\AA, and shows that reprocessed X-rays are just sufficient to
account for the observed 1315\AA\ flux. The 12 eV black body
contributes only 5\% of the flux at 1315\AA\ and demonstrates that any
X-rays reprocessed in NGC 7469 could plausibly be hidden in the EUV
and soft X-ray bands, even if it is strongly peaked.
\label{fig:sed}}
\end{figure}

We now consider the alternative that the X-ray emission drives the UV
in the thermal reprocessing scenario.  The luminosities in the X-ray
and UV bands are similar (Fig.~\ref{fig:sed}) and thermal reprocessing
can therefore be energetically important.  As stated above, the 2-10
keV observed flux of NGC 7469 at our epoch was $3.4\times
10^{-11}$\,erg cm$^{-2}$ s$^{-1}$. However, the X-ray emission of NGC
7469 covers a far wider band than this, with significant emission
being observed down to $\sim 0.1$~keV with ROSAT (Brandt \etal\ 1993)
and most likely up to at least 100~keV as seen by OSSE (Zdziarski
\etal\ 1995; Gondek \etal\ 1996). Estimates of the underlying photon
index of the continuum are in the range $1.9-2.0$ after accounting for
the effects of Compton reflection (Piro \etal\ 1990; Nandra \& Pounds
1994).  We estimate the mean X--ray luminosity of NGC 7469 at our
epoch to be $1.8\times 10^{44}$\,erg s$^{-1}$ in the 0.1-100~keV
band. In the canonical thermal--reprocessing scenario about half of
this luminosity should be absorbed in the accretion disk or other
material. After estimating that fraction which is Compton scattered
rather than absorbed (George \& Fabian 1991) we conclude that the
presence of the iron emission line and reflection hump in NGC 7469
indicate that a luminosity of $\sim8\times 10^{43}$\,erg s$^{-1}$ of
the X-ray emission of NGC 7469 is reprocessed and re-emerges as
thermal emission.

Let us now suppose that all of this luminosity emerges in a single
black body (which represents the narrowest physically-realistic
spectrum) peaking close to 1315\AA\ ($kT \sim 2$~eV).  Such a
blackbody is almost sufficient to account for the continuum at
1315\AA\ (Figure~\ref{fig:sed}).  Therefore, it is energetically
possible that reprocessed X-rays produce some of
the observed UV continuum and
its variations. As in the case of upscattering, the simplest thermal
reprocessing models predict a strong positive correlation between the
bands, with any time lags being in the sense that the X-ray variations
lead those in the UV. No such lag is observed. We do find that the
strongest {\it anti}--correlation of the datasets occurs for the
X-rays leading the UV, but the interpretation of such a result is far
from obvious and we do not comment on it further.

Even if aliasing has caused us to ``miss'' a positive correlation with
a long ($\sim 14$~d) lag between the UV and X--rays, any lag longer
than a day or so is very difficult to explain. Any single transfer
function which related the two bands would smooth the light curve of
the responding band and reduce its amplitude. We observe, however,
that the amplitudes of variability on long time scales are very
similar. We therefore reject such a possibility. In this case it is
even more difficult to envisage a complex geometry which can reproduce
the light curves.

It therefore seems highly unlikely that any substantial proportion of
the 1315\AA\ continuum of NGC 7469 arises from thermal reprocessing
unless, for example, there is substantial anisotropy of the X-ray
emission.  W97 found that the variations at longer UV wavelengths
followed those at shorter wavelengths, but with a time lag of a
fraction of a day.  Collier \etal\ (1998) have demonstrated that this
trend continues into the optical, and Peterson \etal\ (1998) find this
trend to be significant at no less than the 97~per cent confidence
level. This, together with the rapidity of the variations in NGC 7469,
is most easily explicable in terms of the thermal reprocessing
hypothesis. However, our data essentially rule out models in which all
the observed optical/UV flux is re-radiated X-ray emission. The
optical/UV variability therefore requires either intra-band
reprocessing, which is difficult from an energetics standpoint, or
some other model.

Should we therefore conclude that the UV/optical continuum in NGC 7469
arises from direct emission by the accretion disk?  Our data offer no
direct constraints on accretion disk models, as no explicit
relationship between the X--ray and UV emissions is predicted by those
models. Nonetheless, the fact that thermal reprocessing is strongly disfavored
by our data has profound implications for the disk models. The rapid
and wavelength-coherent variations in the optical and UV flux of AGN
is difficult to reconcile with a standard $\alpha$-disk (e.g. Krolik
\etal\ 1991; Molendi, Maraschi \& Stella 1992). 
Prior to our observations, it was conceivable that
thermal reprocessing was responsible for these rapid variations.  There is now
a clear need for a revision of accretion disk theory to account for
these wavelength-independent variations without resorting to
reprocessing.

\subsubsection{The extreme ultraviolet (EUV) continuum and UV emission lines}

Given the presence of a typical iron K$\alpha$ line and reflection
hump in this source (Piro \etal\ 1990; Nandra \& Pounds 1994) we are
left with the question of where the putative reprocessed X-ray flux is
emitted. One possibility is that the thermal reprocessing occurs in a
molecular torus (Ghisellini, Haardt \& Matt 1994; Krolik, Madau \&
Zycki 1994), in which case it might emerge in the infrared. Such an
hypothesis would predict a narrow iron K$\alpha$ line in the X-ray
spectrum, whereas in many Seyfert galaxies these lines are extremely
broad. The case of NGC 7469 is unclear, with Guainazzi \etal\ (1994)
finding no evidence for a broad component and Nandra \etal\ (1997)
finding marginal evidence. A conclusive determination requires a
longer exposure with \asca, but it seems highly likely that the iron
K$\alpha$ line and Compton hump in Seyfert 1 galaxies in general are
produced extremely close to the central black hole (e.g., Nandra
\etal\ 1997). We would therefore expect the reprocessed emission to
emerge at a higher energy. As shown above, however, the thermally
reprocessed X-rays only make a strong contribution to the observed
optical and UV wavebands if the emission is strongly peaked at those
wavelengths. It seems more likely that the emission covers a
range of temperatures, in which case the reprocessed flux would be
difficult to detect when spread over a wide band.  

Alternatively, it could peak in the (unobserved) EUV band.
Figure~\ref{fig:sed} shows that this can indeed be the case. A
blackbody of luminosity $8\times 10^{43}$\,erg s$^{-1}$ contributes
less than 5~per cent of the flux at 1315\,\AA\ as long as $kT>12$~eV.
Intensity variations in such a component would be undetectable with
\iue. Similarly, the \asca\ spectrum constrains $kT<60$~eV. If the
spectral form is broader than a single blackbody, the range of allowed
temperatures is correspondingly wider. Interestingly, Brandt \etal\
(1993) reported evidence for a soft excess in the \ros/PSPC data which
can be modeled as a blackbody of $kT\sim 110$~eV and a luminosity of
$10^{43}$\,erg s$^{-1}$.  This can be identified with the high energy
tail of such a broad, reprocessed component.

\begin{figure}
\plotone{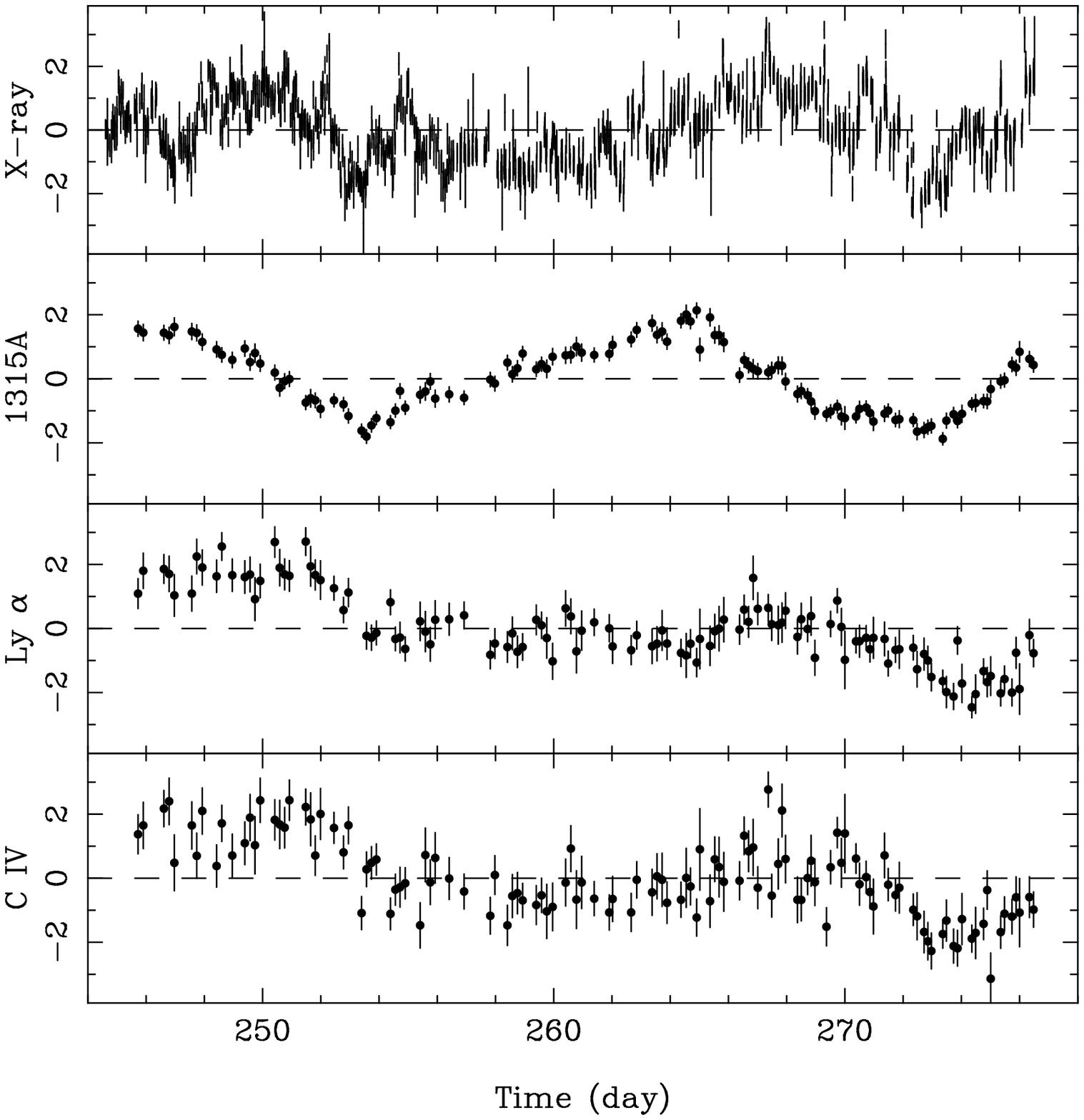}
\caption{
Light curves, renormalized to the rms variability parameter ($F_{\rm var}$)
for (from top to bottom) the 2-10 keV X-rays, the 1315\AA\ continuum,
Ly$\alpha$ and C{\sc iv} emission lines. The emission lines show
a clear long-term fading which is not present in either continuum
band.
\label{fig:renorm_lc}}
\end{figure}

The major UV emission lines are excited by unobservable EUV photons.
An extrapolation of the X-ray spectrum observed by \asca\ (George
\etal\ 1998) and of the UV spectrum into that band indicate roughly
equal contributions at energies at which the lines are excited.  With
the two components being poorly correlated at zero lag, it is
therefore difficult to determine which will be the dominant EUV
component at any given time.  We have suggested above that there may
even be a third contributor to the EUV, the reprocessed X-rays.  In
other words, the shape of the ionizing continuum changes with time.
This effect could account for certain difficulties which have been
encountered in explaining the emission line responses to the observed
UV continuum in reverberation mapping experiments. Our observations
suggest that the unseen EUV continuum is not directly related to the
observed \iue\ flux, which therefore cannot be assumed to be a perfect
representation of the continuum driving the line emission. It is also
interesting to note that the emission-line light curves show long term
trends which are not apparent in the continuum bands. This is most
clearly demonstrated by Fig.~\ref{fig:renorm_lc}, which shows the
X-ray and UV continuum light curves, together with those of the
Ly$\alpha$ and C{\sc iv} emission lines. These have all been
renormalized to the $F_{\rm var}$ value and thus the y-axes crudely
represent the number of standard deviations from the mean. Both
emission lines clearly show a long-term reduction in their flux which
is not seen in either continuum.

\subsection{Steps towards a new model}

In the light of the above, it is clear that new or more complex models
must be sought to explain the data which have been obtained thus far,
and particularly those described in this paper. Here we suggest some
ways in which our new data might be reconciled with the existing
paradigm by modification or extension. We emphasize that such a discussion
is incomplete and {\it ad-hoc}.

As we have stated above, it seems most likely that the X--ray flux
which is absorbed when the iron K$\alpha$ lines is being generated
emerges in a relatively weak, broad component, that may peak the
EUV/soft X-ray band. The emission in this band may well provide the
crucial connection between the higher and lower-energy components.  A
reasonable interpretation of the longer-timescale variability observed
in our light curve is that the UV emission leads that in the X-rays,
but with a variable lag. This suggests the dominant source of
variations is in the seed population of an upscattering model. We do,
however, bear in mind the caveat that the particle distribution of the
upscattering medium must also be variable, to produce the most rapid
variations. To explain the ``variable'' time lag, we suggest that
there are multiple ``seed'' populations, which dominate at different
times. In particular we suggest that the main source of 1315\AA\
photons is located at a distance of $\sim 4$~lt d from the X--ray
source and they are the dominant seed population when the source is in
a high flux state, thus introducing a 4d ``lag'' between the X-rays
and UV. When the 1315\AA\ flux is observed to decline, however, this
allows other emitting regions to dominate the seed distribution. In
particular we suggest at these times that EUV/soft X-ray photons are
the dominant seed population. They arise from closer in and are
therefore observed to have little or no lag with the X-rays.

As might be apparent from the above discussion, the primary X--rays,
reprocessed X-rays and primary UV might well exist in a rather fine
balance in the typical AGN. Future observational data on other objects
of similar quality to that presented here, and preferably including
the far-UV and soft X-ray bands, will be necessary for further
progress and to establish the generality of the phenomena explored
here.

\acknowledgements

We thank the \xte\ team for their operation of the satellite, the
\xte\ GOF for their assistance in data analysis, Keith Jahoda and Mike
Stark for their extensive work on the background subtraction. We also
thank Tal Alexander and Shai Kaspi for assistance with the ZDCF, Hagai
Netzer and Iossif Papadakis for discussions, and the anonymous referee
for many helpful suggestions.  We acknowledge the financial support of
the National Research Council (KN) and Universities Space Research
Association (IMG,TJT).  We are grateful for the support of this work
through NASA grants NAG5-3233 and NAG5-3497 to Ohio State University.
This research has made use of the Simbad database, operated at CDS,
Strasbourg, France; of the NASA/IPAC Extragalactic database (NED)
which is operated by the Jet Propulsion Laboratory, Caltech, under
contract with NASA; and of data obtained through the HEASARC on-line
service, provided by NASA/GSFC.

\clearpage

\begin{deluxetable}{lllllll}

\tablecolumns{2}
\tablecaption{\xte\ selection criteria
\label{tab:sel}}

\tablehead{
\colhead{Criterion} & \colhead{Description}
}
\startdata
{\tt OFFSET < 0.01} & Angular offset from nominal pointing position 
  (\arcdeg) \nl 
{\tt ELV > 10} & Angle from earth's limb (\arcdeg) \nl
{\tt NUM\_PCU\_ON > 2} & Number of operational PCUs \nl
{\tt (VpX1L+VpX1R)/Q6 < 0.1 } & Normalized propane layer veto rate \nl

\tablecomments{Only data from PCUs 0,1 and 2 were employed; VpX1L
represents the veto rate between the PCA propane layer and the left
half of the Xenon layer; VpX1R is for the right half of the Xenon
layer; Q6 represents the veto rate between 6 of the 8 anode chains in
the PCA}

\enddata
\end{deluxetable}

\begin{deluxetable}{lllllll}

\tablecolumns{7}
\tablecaption{Variability parameters
\label{tab:var}}

\tablehead{
\colhead{Band} & \colhead{$N_{\rm data}$} &
\colhead{$\overline{F}$} & \colhead{$\sigma_{\rm F}$} &
\colhead{$\Delta_{\rm F}$} & \colhead{$\sigma_{\rm XS}$} &
\colhead{$F_{\rm var}$} 
}
\startdata
2-10 keV    & 1513 & 8.59  & 1.41 & 0.47 & 1.33 & 0.155 \nl
1315 \AA    & 123  & 4.38  & 0.66 & 0.16 & 0.65 & 0.147 \nl
Ly$\alpha$  & 121  & 203.3 & 17.5 & 7.99 & 15.5 & 0.076 \nl
N {\sc v}   & 123  & 38.5  & 7.04 & 6.22 & 3.28 & 0.085 \nl
Si {\sc iv} & 123  & 54.9  & 11.4 & 9.60 & 6.13 & 0.112 \nl
C  {\sc iv} & 123  & 271.9 & 19.0 & 10.6 & 15.7 & 0.057 \nl
He {\sc ii} & 123  & 65.9  & 13.6 & 13.2 & 3.39 & 0.051 \nl
C {\sc iii} & 123  & 60.5  & 8.94 & 15.1 & \nodata & \nodata \nl

\tablecomments{The UV light curves have been truncated to cover the
same baseline as the X-ray light curve. $N_{\rm data}$ is the number
of data points; $\overline{F}$ is the mean flux in ct s$^{-1}$ for the
X-ray and in units of $10^{-14}$\,erg cm$^{-2}$ s$^{-1}$ \AA$^{-1}$ for
the UV; $\sigma_{\rm F}$ is the square root of the variance of the
light curve. $\Delta_{\rm F}$ is the mean error bar; $\sigma_{\rm XS}$
is the square root of the excess variance and $F_{\rm var}$ is the
square root of the normalized excess variance, which we term the rms
variability parameter (see text)}

\enddata
\end{deluxetable}

\begin{deluxetable}{llll}
\small

\tablecolumns{4}
\tablecaption{Correlation coefficients  \label{tab:ccf}}

\tablehead{
\colhead{Band} & \colhead{$r_{\rm 0}$} & 
\colhead{$r^{+}_{\rm max}$} & \colhead{$r^{-}_{\rm max}$}
}

\startdata
\cutinhead{Correlations with the X-ray light curve}
1315 \AA    & 0.19 & 0.45 & -0.57 \nl
Ly$\alpha$  & 0.28 & 0.35 & -0.35 \nl
N {\sc v}   & 0.24 & 0.26 & -0.24 \nl
Si {\sc iv} & 0.22 & 0.23 & -0.37 \nl
C  {\sc iv} & 0.30 & 0.39 & -0.33 \nl
He {\sc ii} & 0.14 & 0.18 & -0.24 \nl
\cutinhead{Correlations with the truncated UV light curve}
2-10 keV    & 0.18 & 0.46 & -0.59 \nl
Ly$\alpha$  & 0.25 & 0.56 & -0.32 \nl
N {\sc v}   & 0.18 & 0.41 & -0.21 \nl
Si {\sc iv} & 0.35 & 0.53 & -0.29 \nl
C  {\sc iv} & 0.20 & 0.49 & -0.36 \nl
He {\sc ii} & 0.38 & 0.42 & -0.30 \nl
\tablecomments{
$r_{\rm 0}$ is the correlation coefficient between X-ray or
truncated UV and the
other component at zero lag; 
$r^{+}_{\rm max}$ is the maximum ({\it positive}) correlation coefficient
in the range $\pm 10$~d;
$r^{-}_{\rm max}$ is the minimum ({\it negative}) correlation coefficient
in that range
}
\enddata
\end{deluxetable}

%-----------------------------------------------------------------------
%\newpage

\clearpage

\clearpage

\clearpage

\clearpage

\clearpage

\clearpage

\clearpage


\clearpage
\begin{references}

\reference{ab91} Abramowicz, M.A., Bao, G., Lanza, A., Zhang, X.-H., 1991, 
                 A\&A, 245, 454
\reference{al97} Alexander, T., 1997, in ``Astronomical Time 
                 Series'', D. Maoz, A. Sternberg, E. Leibowitz, Eds, 
                 Kluwer, Dordrecht, p. 163
\reference{bm82} Blandford, R.D., McKee, C.F., 1982, ApJ, 255, 419
\reference{br93} Brandt, W.N., Fabian, A.C., Nandra, K., Tsuruta, S.,
                 1993, MNRAS, 265, 996
\reference{ch85} Chapman, G.N.F., Geller, M.J., Huchra, J.P., 1985,
	         ApJ, 297, 151
\reference{cl91} Clavel, J., \etal, 1991, ApJ, 366, 64
\reference{cl92} Clavel, J., \etal, 1992, ApJ, 393, 113
\reference{co98} Collier, S., \etal\ 1998, ApJ, in press
\reference{co90} Courvoisier, T. J.-L., \etal, 1990, A\&A, 234, 73 
\reference{cc91} Courvoisier, T., J.-L., Clavel, J., 1991, A\&A, 248, 389
\reference{do90} Done, C., Ward, M.J., Fabian, A.C., Kunieda, H.,
                 Tsuruta, S., Lawrence, A., Smith, M.G., Wamsteker, W.,
	         1990, MNRAS, 243, 713
\reference{ek88} Edelson, R.A., Krolik, J.H., 1987, ApJ, 333, 646
\reference{em86} Edelson, R.A., Malkan, M.A., 1986, ApJ, 308, 59
\reference{ed90} Edelson, R.A., Pike, G.F., Krolik, J.H., 
                 1990, ApJ, 359, 86
\reference{ed96} Edelson, R.A., \etal, 1996, ApJ, 470, 364
\reference{fa95} Fabian, A.C., Nandra, K., Reynolds, C.S., Brandt, W.N.,
                 Otani, C., Tanaka, Y., Inoue, H., Iwasawa, K., 1995,
                 MNRAS, 277, L11
\reference{gp87} Gaskell, C.M., Peterson, B.M., 1987, ApJS, 65, 1
\reference{ge95} Genzel, R., Weitzel, L., Tacconi-Garman, L.E.,
                 Blietz, M., Cameron, M., Krabbe, A., Lutz, D.,
                 Sternberg, A., 1995, ApJ, 444, 129
\reference{ge91} George, I.M., Fabian, A.C., 1991, \mnras, 249, 352
\reference{ge98} George, I.M., Turner, T.J., Netzer, H., Nandra, K.,
                 Mushotzky, R.F., Yaqoob, T., 1998, ApJS, 114, 73
\reference{gh93} Ghisellini, G., Haardt, F., Matt, G., 1994,
                 MNRAS, 267, 743
\reference{go96} Gondek, D., Zdziarski, A.A., Johnson, W.N., George, I.M.,
                 McNaron-Brown, K., Magdziarz, P., Smith, D., Gruber, D.E.,
                 1996, MNRAS, 282, 646
\reference{gr93} Green, A.R., McHardy, I.M., Lehto, H.J., 1993, MNRAS, 
                 265, 664
\reference{gu94} Guainazzi, M., Matsuoka, M., Piro, L., Mihara, T.,
                 Yamauchi, M., 1994, ApJ, 436, L35
\reference{gu83} Guilbert, P.W., Fabian, A.C., Rees, M.J., 1983, MNRAS,
                 205, 593
\reference{gr88} Guilbert, P.W., Rees, M.J.,
	         1988, MNRAS, 233, 475
\reference{hm91} Haardt, F., Maraschi, L., 1991, ApJ, 380, 51
\reference{hm93} Haardt, F., Maraschi, L., 1993, ApJ, 413, 507
\reference{ha94} Haardt, F., Maraschi, L., Ghisellini, G., 1994, ApJ, 432, L95
\reference{ha97} Haardt, F., Maraschi, L., Ghisellini, G., 1997, ApJ, 476, 620
\reference{ja96} Jahoda, K., Swank, J.H., Giles, A.B., Stark, M.J., 
                 Strohmayer, T., Zhang, W., Morgan, E.H., 1996,
                 EUV, X-ray and Gamma-ray Instrumentation for Space Astronomy 
                 VII, O. H. W. Siegmund and M. A. Grummin, Eds.,
                 SPIE 2808, p. 59
\reference{kr80} Kriss, G.A., Canizares, C.R., Ricker, G.R., 1980,
                 ApJ, 242, 492
\reference{kr91} Krolik, J.H., Horne, K., Kallman, T.R., Malkan, M.A.,
                 Edelson, R.A., Kriss, G.A., 1991, ApJ, 371, 541
\reference{kr94} Krolik, J.H., Madau, P., Zycki, P.T., 1994, ApJ, 420, L57
\reference{lp93} Lawrence, A., Papadakis, I., 1993, ApJ, 414, L85
\reference{li88} Lightman, A.P., White, T.R., 1988, \apj, 335, 57
\reference{ma82} Malkan, M.A., Sargent, W.L., 1982, \apj, 254, 22
\reference{mn89} Maoz, D., Netzer, H., 1989, MNRAS, 236, 21
\reference{mr91} Maraschi, L., Chiappetti, L., Falomo, R., Garilli, B.,
                 Malkan, M.A., Tagliaferri, G., Tanzi, E.G., Treves, A.,
                 1991, ApJ, 368, 138
\reference{ma81} Marshall, N., Warwick, R.S., Pounds, K.A., 1981, MNRAS, 
                 194, 987
\reference{ma91} Matt, G., Perola, G.C., Piro, L., 1991, \aap, 245, 63
\reference{mo92} Molendi, S., Maraschi, L., Stella, L., 1992, MNRAS, 255, 27 
\reference{mo86} Morini, M., \etal, 1986, ApJ, 307, 486
\reference{mu93} Mushotzky, R.F., Done, C., Pounds, K.A., 1993,
		 ARAA, 31, 717
\reference{na94} Nandra, K., Pounds, K.A., 1994, \mnras, 268, 405
\reference{na97} Nandra, K., George, I.M., Mushotzky, R.F., Turner, T.J.,
                 Yaqoob, T., 1997, ApJ, 477, 602
\reference{np97} Netzer, H., Peterson, B.M., 1997, in Astronomical Time
                 Series, ed.\ D. Maoz, A. Sternberg, E.M. Leibowitz 
                 (Dordrecht: Kluwer), p.\ 85
\reference{pl93} Papadakis, I., Lawrence, A., 1993, MNRAS, 261, 612
\reference{pc95} Perez-Olea, D.E., Colina, L., 1995, ApJ, 468, 191,  
\reference{pe86} Perola, G.C., \etal, 1986, ApJ, 306, 508 
\reference{pe93} Peterson, B.M., 1993, PASP, 105, 247
\reference{pe91} Peterson, B.M., \etal, 1991, ApJ, 368, 119
\reference{pe98} Peterson, B.M., Wanders, I., Horne, K., Collier, S.,
                 Alexander, T., Kaspi, S., Maoz, D., 1998, PASP, in press
\reference{pi90} Piro, L., Yamauchi, M., Matsuoka, M., 1990, ApJ, 360, L35
\reference{pr92} Press, W.H., Teukolsky, S.A., Vetterling, W.T.,
                 Flannery, B.P., 1992, {\it Numerical Recipes},
                 Cambridge University Press, Cambridge, UK
\reference{ro97} Rodriguez-Pascual, P.M., \etal, 1997, ApJS, 110, 9
\reference{sl95} Santos-Lle\'{o}, M., Clavel, J., Barr, P., Glass, I.S., 
                 Pelat, D., Peterson, B.M., Reichert, G.A., 1995, 
                 MNRAS, 274, 1
\reference{ss72} Shakura, N.I., Sunyaev, R.A., 1973, A\&A, 24, 337 
\reference{sh78} Shields, G.A, 1978, \nat, 272, 706
\reference{st95} Stern, B.E., Poutanen, J., Svensson, R., 
                 Sikora, M., Begelman, M.C., 1995, ApJ, 449, L13
\reference{sm89} Sun, W.-H., Malkan, M.A., 1989, ApJ, 346, 68
\reference{st80} Sunyaev, R.A., Titarchuk, L.G., 1980, A\&A, 86, 121
\reference{sv83} Svensson, R., 1983, ApJ, 270, 300
\reference{ta95} Tanaka, Y., \etal, 1995, \nat, 375, 659
\reference{tr90} Treves, A., Bonelli, G., Chiappetti, L., Falomo, R., 
                 Maraschi, L., Tagliaferri, G., Tanzi, E.G., 1990,
                 ApJ, 359, 98
\reference{wa97} Wanders, I., \etal, 1997, ApJS, 113, 69
\reference{wp94} White, R.J., Peterson, B.M., 1994, PASP, 106, 879
\reference{zd94} Zdziarski, A.A., Fabian, A.C., Nandra, K., Celotti, A.,
                 Rees, M.J., Done, C., Coppi, P.S., Madejski, G.M., 1994,
                 MNRAS< 269, L55
\reference{zd90} Zdziarski, A.A., Ghisellini, G.G., George, I.M.,
                 Fabian, A.C., Svensson, R., Done, C., 1990, ApJ, 363, L1
\reference{zd95} Zdziarski, A.A., Johnson, W.N., Done, C., Smith, D.,
                 McNaron-Brown, K., 1995, ApJ, 438, L63

\end{references}
\end{document}